# The Minimal Model of Financial Complexity

**Philip Z. Maymin**
NYU-Polytechnic Institute
Department of Finance and Risk Engineering
Six MetroTech Center
Brooklyn, NY 11201
Phone: (718)260-3175
Fax: (718)260-3355
Email: phil@maymin.com

## *Abstract*

A representative investor generates realistic and complex security price paths by following this trading strategy: if, a few ticks ago, the market asset had two consecutive upticks or two consecutive downticks, then sell, and otherwise buy. This simple, unique, and robust model is the smallest possible deterministic model of financial complexity, and its generalization leads to complex variety. Compared to a random walk, the minimal model generates time series with fatter tails and more frequent crashes, thus more closely matching the real world. It does all this without any parameter fitting.

## 1. Introduction

What is the simplest rule to model the complexity and randomness of financial markets? Security prices have complex behavior in the time series and cross section of prices, returns, volume, and liquidity. It's easy to mimic security prices by introducing randomness, either externally as with a random walk (see e.g. Malkiel 2003) or internally through large structures such as cellular automata involving trade between many investors (e.g. Wolfram 2002, page 432), but what are the *minimal* models that generate complexity?

I find that a single investor trading in a single security can generate complex price behavior all by himself. He makes trading decisions by looking back at the signs of the past few price changes, which can be interpreted as days or seconds or even ticks. Because he is the representative investor, the price adjusts with his decisions. This simple framework allows for an infinite variety of complex price series. An online demonstration (Maymin *2007a*) allows interactive exploration of this framework.

In particular, I identify the minimal model in this framework, determine that it is essentially unique and represents a simple delayed-trading strategy, and show that it is robust in that in generates complexity for many different values of the lookback parameter.

The statistical properties of the minimal model relative to a random walk—higher kurtosis and a more negative skew—also more closely match broadly observed empirical phenomenon of the real world. As we will see, there are only two parameters, the rule number corresponding to the minimal model and the lookback window, and they cannot be tweaked to "fit" the results in the traditional sense. Of the 256 two-state, buy-sell rules, only one consistently generates complex behavior, and it does so for a variety of lookback windows.

The minimal model therefore serendipitously seems to generate complex price series that resemble reality.

## 2. Walkthrough

Let's start with a walkthrough example. Our representative investor has a three-day lookback window, meaning he only cares about the direction the market took over each of the past three days. He comes into work on Thursday and sees the following history for the market: **Monday was down, Tuesday was up, and Wednesday was up**.

Our trader is modeled by the following rule, which we will come to know as rule 54.

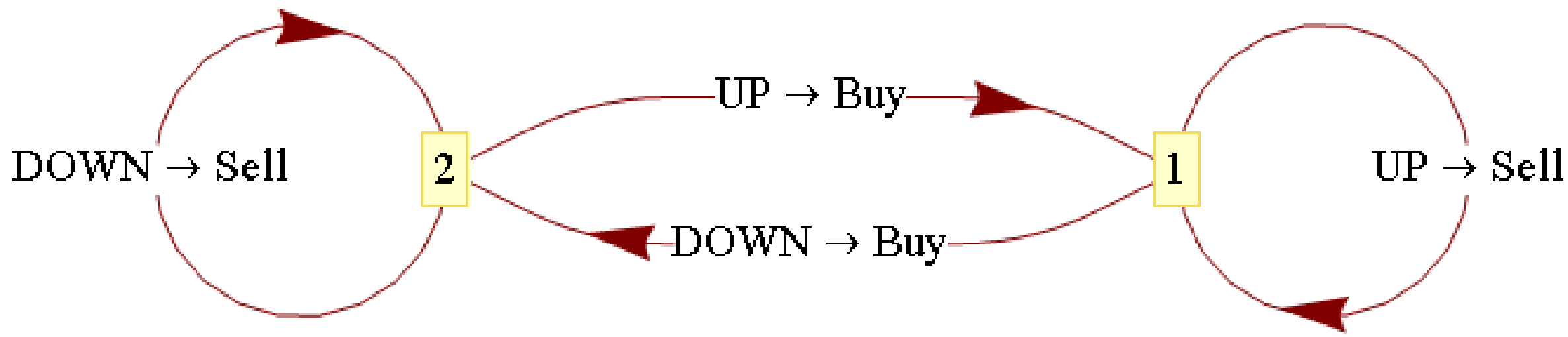

**Figure 1: Transition Diagram of Rule 54.** The investor begins in state 1 and looks back on past movements to determine whether to buy or sell.

Here is how to understand the above rule. The two boxes labeled 1 and 2 represent mental states of the investor during his decision process. You can loosely think of them as emotional states like "happy" or "jealous" or "frightened," though they are more specific than that. The investor always starts in state 1 at the beginning of each new day.

Two directed arrows leave each of the two states, one for UP and one for DOWN. On Thursday, the investor, starting in state 1, sees that the market was UP on the previous day, Wednesday. So he follows the UP→Sell arrow out of state 1, which returns him back to the same state.

What does the "→Sell" portion after UP mean on the directed arrow? It is his current thinking; it means that if the trader were to stop looking at price histories at this point, he would be inclined to sell. In other words, our trader, having looked back at one day's price change and seen that yesterday the market was up, is bearish.

But he doesn't trade yet. He continues looking back. When we left him, he had just followed the UP arrow to remain in state 1. He now notes that the day before yesterday, Tuesday, was also up, and again follows the same arrow, again remaining in state 1.

But now he looks at Monday, which was down. Now he follows the DOWN→Buy arrow out of state 1 and into state 2. If his lookback window exceeded three, he would continue looking at even earlier days, following the relevant UP or DOWN arrows across his various states.

In our case, he has reached his limit. The last arrow he followed was DOWN→Buy, so he will buy the market asset.

Because he is the representative investor, the market rises on Thursday.

Assuming a single, representative investor allows us to model with just one agent a market that by definition is a history of traded prices between multiple parties. In contrast to multiple agent models such as e.g. Gonçalves (2003) or Zhou and Sornette (2007), we are searching for the minimal model of complexity, and it seems that much of the complexity can still be generated even with just one investor and just one asset.

With any representative investor model, no trading actually takes place because there is no one for him to trade with. Instead, the market price merely adjusts until the representative investor is indifferent.

However, our trader does not know this. From his point of view, he put in an order to buy and was simply unlucky enough not to have been filled before the market moved away. This happens to him every day, whether he wants to buy or sell, and he doesn't learn from it. He also never learns his own rule. This is a very simple model.

We will see later that rule 54 is in fact the simplest possible lookback model to generate a complex, realistic-looking price series. In principle, the investor could use a rule with more states and more possible actions.

On Friday, the trader comes in to work and sees the following history for the market: **Tuesday was up, Wednesday was up, and Thursday was up**. It is easy to see that he will now want to sell, so the market will go down. The process repeats every day.

An online demonstration (Maymin 2007*b*) allows an interactive walk-through for this and other rules and initial conditions.

## 3. Formal Model

An investor with $s$ internal states and $k$ possible "actions" with base $b$ and having a lookback window of $w$ days is modeled as an iterated finite automaton (IFA).

What are actions? When $k = 2$, the possible actions are buy and sell. When $k = 3$, the possible actions are buy, sell, and hold. When $k = 4$, the possible actions are buy, buy more, sell, and sell more. In general, if $k$ is even, the possible actions are $k/2$ types of buys and $k/2$ types of sells, with differing "strengths." If $k$ is odd, the possible actions are the same as for $k - 1$ with the additional possibility of doing nothing, i.e., hold.

The "strength" of a buy or sell is an exponent of the base $b$. For example, when $b = 3$ and $k = 4$, the four possible actions are -3, -1, 1, and 3, meaning sell three times more fervently than normal, sell normally, buy normally, or buy three times more fervently than normal. If $k = 6$, then two additional actions are -9 and 9.

What does it mean to buy or sell more fervently? In the context of a representative investor, it is a scaling factor. If a normal buy would increase the market price by one percent, a nine-times more fervent order would increase the market price by nine percent. In principle, the base $b$ should depend on the liquidity of the underlying security. (A more complicated model with multiple agents could allow $b$ to vary depending on the liquidity available at the time.)

The walkthrough of section 2 describes a representative investor with $s = 2$, $k = 2$, and $w = 3$. Because $k < 4$, the base $b$ is irrelevant.

Wolfram (2003) defines iterated finite automata and provides a numbering scheme to uniquely identify an IFA for a given $s$ and $k$:

> Any finite automaton can be represented by a network, in which each node is a state, and each edge represents a transition from one state to another... A transducer finite automaton, however, is set up to take in one sequence of symbols, and put out another. The way it works is to have both an input and an output symbol on each edge in the network. Transducer finite automata are sometimes known as Mealy machines, and particularly in the past, they were often used as models of electrical or mechanical machines that operate in a sequential way.

A convenient way to represent a transducer finite automaton is to specify each edge in its network by giving a rule of the form {state1, input} -> {state2, output}...

There are (k s)^(k s) s-state, k-symbol transducer finite automata—or 256 2-state, 2-symbol ones. One can number finite automata in analogy to the numbering of Turing machines from page 888R [of Wolfram (2002)]. The rule corresponding to a finite automaton with number m is then [in Mathematica code]:

```
ToFARule[m_Integer, {s_Integer, k_Integer}] :=
   Flatten[MapIndexed[{1, -1}#2 + {0, k} ->
   Mod[Quotient[#1, {k, 1}], {s, k}] + {1, 0} &,
   Partition[IntegerDigits[m, s k, s k], k], {2}]]
```

By Wolfram's convenient IFA numbering scheme, the IFA of figure 1 is for rule 54. Running the code above for $m = 54$, $s = 2$, and $k = 2$ results in the rules {{1,1}→{1,0}, {1,0}→{2,1}, {2,1}→{1,1}, {2,0}→{2,0}}. The first part of each pair is the state. In the second part of each pair, we interpret 1's as ups and buys and 0's as downs and sells.

The final difference from the standard use of an IFA is that we only look at the final output to determine the trader's desired action, rather than tracking the entire sequence of buy and sell mindsets that the trader fluctuates between while deciding.

The trader cannot make his first decision until there are at least $w$ past days for him to consider. Thus, the initial price history for the first $w$ days can be set arbitrarily. For convenience, it is set by default to be a sequence of all UP days. We will see that for rules that generate complex behavior, this choice is innocuous.

## 3.1. Complexity

But first we need to define complexity. In general, complexity is hard to define, but for our purposes, we can be precise. First note that with a lookback window of $w$ with $k$ possible market movements each day, there are only $k^w$ distinct possible price histories on which the trader makes his decision. Therefore, after $k^w$ days of history, the time series of price changes must have cycled. A complex series is one that takes a long time to cycle. A simple series cycles quickly.

For example, a rule that always buys will cycle in a single day. Even though the price continues to rise to a unique level each day, the changes are constant.[†]

[†] Price changes mean simply the difference in price between successive days. All the results continue to hold in their essence if price changes are replaced with price returns or even log-returns, so for clarity I use only price changes in the text.

**Definition 1**. Complexity in a $k$-action generated price series for a lookback window $w$ means that the periodicity of the rule is greater than $k^w/2$.

A rule with the maximal complexity would have a periodicity equal to the maximal periodicity of $k^w$. However, we relax the definition of complexity mildly to allow for very, though not maximally, complex rules, so long as its cycle is at least half of the maximum.

We will also assume that $w \geq 5$ to avoid degenerate situations where a series is technically complex but only has periodicity of a few days.

Cycling does not mean that the market is modeled as repeating; rather, this is just a characteristic of the model. We can always choose sufficiently large $k$ and $w$ such that the cycle length is longer than the history of the market under consideration (for example, $k = 15$ and $w = 15$ means the maximal cycle length is longer than the age of the universe in seconds).

Our precise definition of complexity lets us quickly establish that the minimal complex model must have at least two states and two actions.

**Lemma 1**. Complexity in the price series requires $s \geq 2$ and $k \geq 2$.

**Proof.** If $k = 1$, there is only one possible action, regardless of the state or price history. Therefore the price series will simply trend.

If $k = 2$ and $s = 1$, there is only one state, so only the direction of the market movement on the earliest day considered matters. There are just four possible rules. Either the rules always buy or always sell, so that the price series simply trends as with $k = 1$; or the rules continue the pattern, buying on UP days and selling on DOWN days, so that the movement on day $t$ equals the movement on day $t - w$ and the rule will have periodicity $w$, far less than the critical periodicty $2^{w-1}$, and hence not complex; or the rules flip the pattern, selling on DOWN days and buying on UP days, so that the movement on day $t$ is the opposite to that of day $t - w$, thus the rule has periodicity $2w$, again far less than the critical periodicity $2^{w-1}$. ■

We can now see why the initial arbitrary price history for the first $w$ days does not matter for complex series. For maximally complex series, their periodicity equals the maximal periodicity. Therefore, every single possible $w$-day history of price changes will be represented on the generated cycle. Thus, starting with a history of all UP days, as opposed to a history of all DOWN days or a mixed history, merely changes the starting point and does not affect the periodicity.

For other complex series, there may be other starting points that are off of the main cycle, but we know that those secondary cycles will have a maximal possible cycle length of less than $k^w/2$ and so will not be complex. In other words, for series that are already complex, we are not missing anything by starting with a history of all UP days.

Finally, for a series that appears simple, it is possible that a different starting history would lead to a complex time series. However, it is unlikely to be a complex series at all, as the chances that the single price history of all UP days is off of the main cycle is $1/k^w$, which becomes smaller for larger $k$ and $w$. And it would surely not be a maximally complex series, because in a maximally complex series, all histories are on the main cycle.

These ideas are illustrated in figure 2, which shows the transition diagrams among all possible price histories. Note that rule 54 has nearly maximal complexity and periodicity when $w = 5, 6, 7$, or 10, but has multiple smaller, non-complex cycles when $w = 8$ or 9. Other rules have some temporary fluctuations but tend to collapse onto a single price history or short cycle. Note also the robustness of the essential features of each of the representative rules across increasing lookback windows: each rule seems to have its own persistent character.

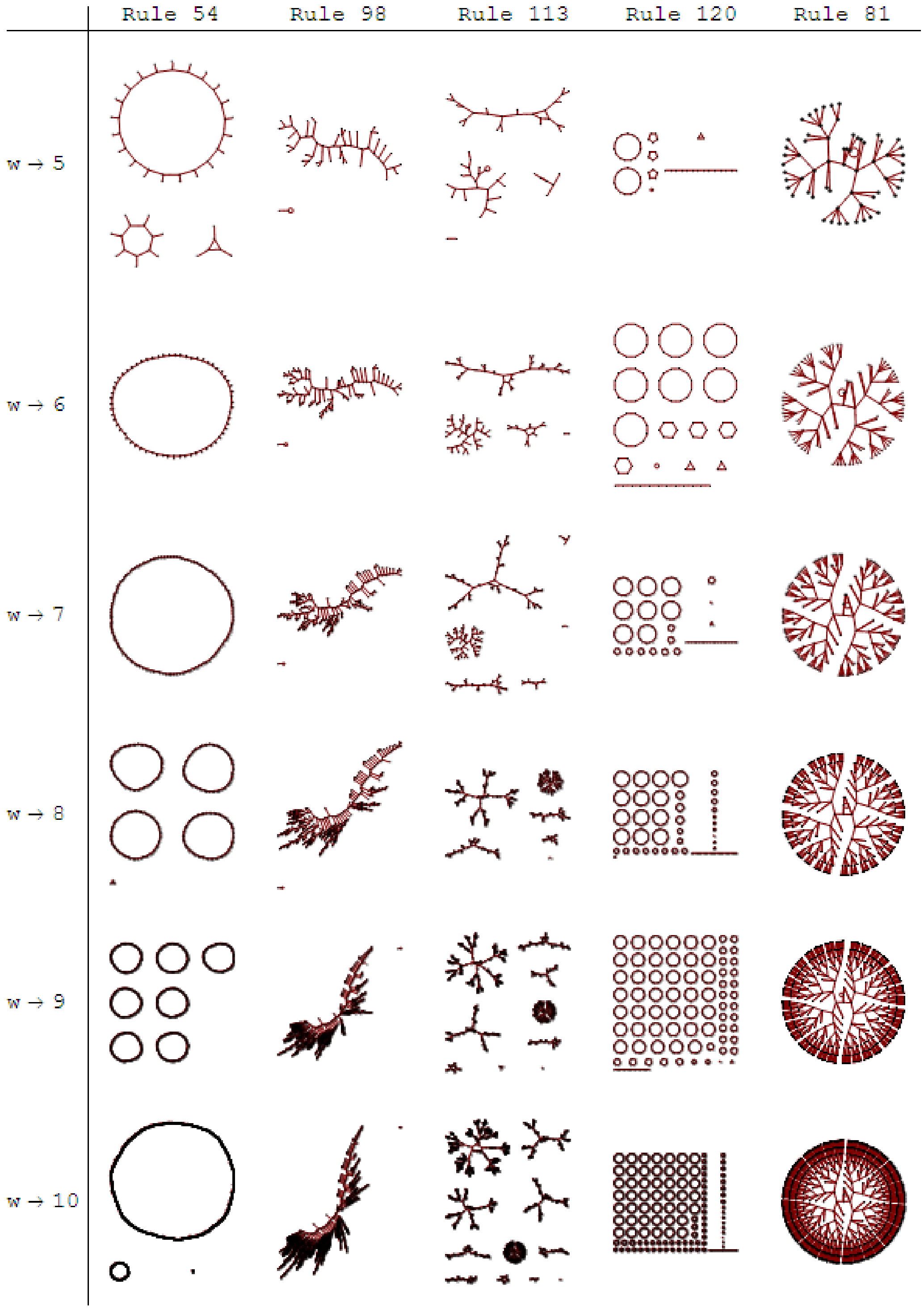

**Figure 2: Transition Diagrams.** Each point in the graph plots below represent a possible sequence of ups and downs of length $w$ for two-state buy-sell rules. The edges between the points represent transitions between histories. Long cycles suggest complexity, such as for rule 54 with lookback window $w = 5$. The other rules were chosen as representative of the variety across the different rules and to show that their essential features also are stable across increasing lookback windows.

Most rules generate only simple behavior. For example, for $s = 3$, $k = 2$, and $w = 9$, only 270 out of the $(2 \cdot 3)^{2 \cdot 3} = 46{,}656$ possible rules generate complex behavior. The first 50 of these are shown in figure 3. Note the wide variety: each of them are complex in their own way.

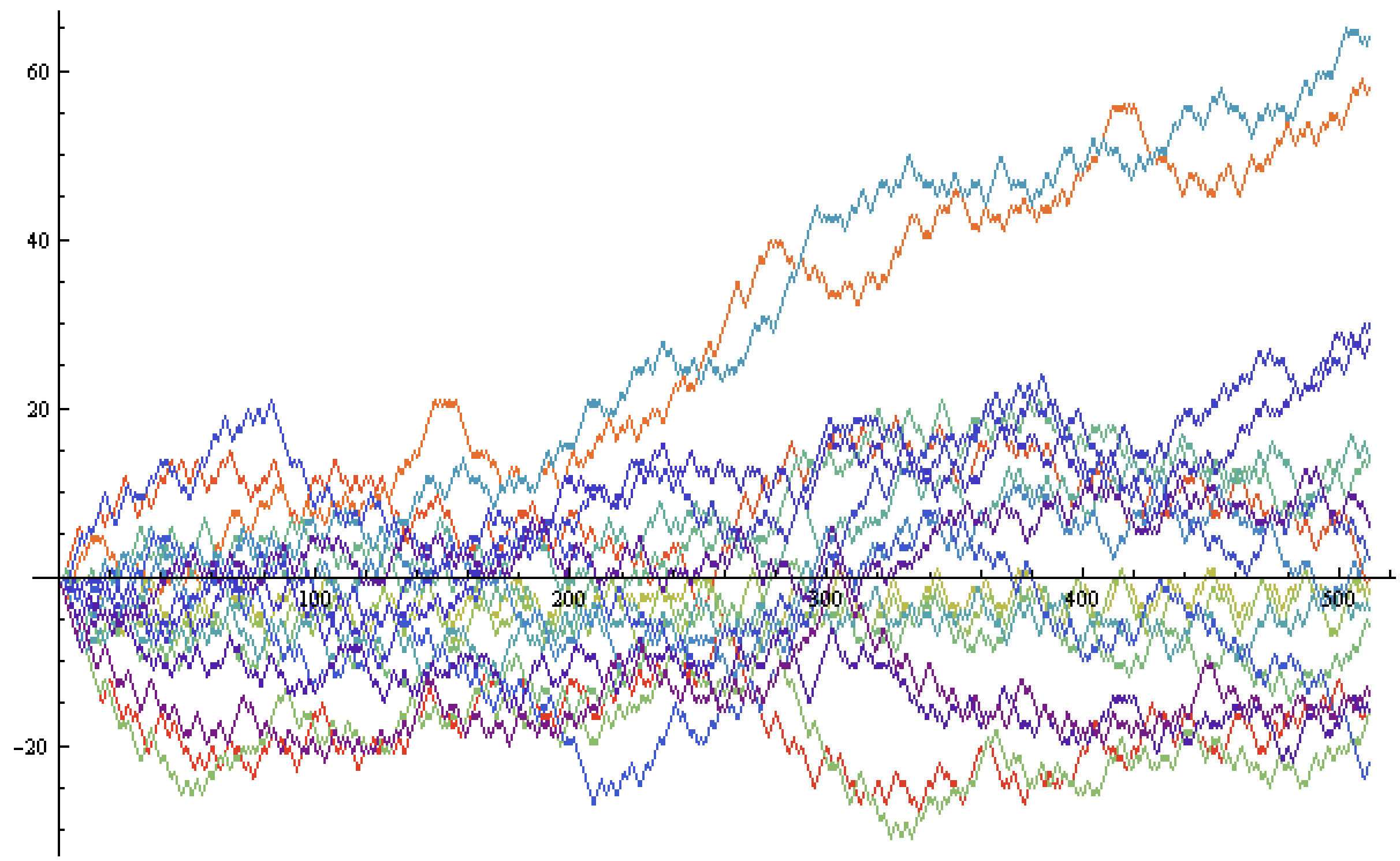

**Figure 3: Some Complex Time Series.** The graph below charts the first $2^9 = 512$ time steps of 50 of the 270 three-state, two-action rules that generate complex behavior with a lookback window of 9 days.

## 4. The Minimal Model

Of the 256 two-state, buy-sell rules, only[‡] rule 54 generates complex behavior with lookback windows ranging from 5 days to 15 days. You can check this for yourself with the online demonstration by looking at different rule numbers and different lookback windows.

What does rule 54 compute? In general there is no easy way to tell what an arbitrary *s*-state, *k*-action automaton will do without simply running it and finding out. In this case, however, there is a shortcut.

‡ Rule 201 is identical to rule 54 except state 1 is labeled state 2, and vice versa, meaning the only difference is, because of our default in always starting in the state called "1," that rule 54 starts in state 1 and rule 201 starts in what rule 54 calls "2." Because of the way these rules work, so long as the lookback window is greater than a single day, they will always have the same output. The reason is that the first day's movement will cause both to go to the state that rule 54 calls "1" on an UP move and will cause both to go to the state that rule 54 calls "2" on a DOWN move. After that, they have become identical.

Observe that state 1 is an UP-absorbing state: any UP day will immediately result in the investor entering state 1. Similarly, state 2 is a DOWN-absorbing state. Therefore, all of rule 54 boils down to the earliest two days of its lookback window.

If they were both up, then the investor's final arrow would be the UP→Sell one out of state 1, and he would sell. If they were both down, then the investor's final arrow would be the DOWN→Sell out of state 2, and he would also sell. If they were up and down, or down and up, then the investor would buy.

In short, rule 54 with a lookback window $w$ is identical to comparing the market movement of $w$ days ago with that of the movement of $w - 1$ days ago: buy if they are different, sell if they are the same. Repeat each day afresh.

An alternative non-lookback interpretation of the minimal model of rule 54 is as follows, interpreting each "day" as only a tick. An investor observes two consecutive ticks. If they are the same sign, i.e., either both up or both down, then he sells. Otherwise he buys. However, his order does not take effect for $w - 1$ ticks. Put another way, his orders face a delay of $w - 1$ ticks. Ticks are approximately on the order of seconds, so the minimal model formalizes an investor who looks at consecutive ticks to decide his position and faces a short delay either in mental processing or order execution.

In either interpretation, the minimal model is robust to the choice of $w$. Recalling that complexity is defined as having a cycle length near the maximal cycle length, table 1 lists the lookback windows (or order delays) $w$ for which rule 54 generated a complex financial price series.

**Table 1: Rule 54 Periods.** This table lists the period (i.e., cycle length) of the series generated by rule 54 for eighteen different lookback windows $w$ from $w = 5$ to $w = 22$. Thirteen of them, or nearly three-quarters, generate complex series, having a period more than half of the maximum period $2^w$. Those rows are shaded in gray. In particular $w = 6$, 7, 15, and 22 are just one iteration short of maximal complexity.

| Lookback | Rule 54 Period | Max Period | % of Max |
|---|---|---|---|
| 5 | 21 | 32 | 65% |
| 6 | 63 | 64 | 98% |
| 7 | 127 | 128 | 99% |
| 8 | 63 | 256 | 24% |
| 9 | 73 | 512 | 14% |
| 10 | 889 | 1,024 | 86% |
| 11 | 1,533 | 2,048 | 74% |
| 12 | 3,255 | 4,096 | 79% |
| 13 | 7,905 | 8,192 | 96% |
| 14 | 11,811 | 16,384 | 72% |
| 15 | 32,767 | 32,768 | 99% |
| 16 | 255 | 65,536 | 0% |
| 17 | 273 | 131,072 | 0% |
| 18 | 253,921 | 262,144 | 96% |
| 19 | 413,385 | 524,288 | 78% |
| 20 | 761,763 | 1,048,576 | 72% |
| 21 | 5,461 | 2,097,152 | 0% |
| 22 | 4,194,303 | 4,194,304 | 99% |

For *w* between 5 and 22, only the five values of *w* = 8, 9, 16, 17, and 21 do not generate a complex price series. The remaining 13 values generate complexity. In particular, the four values of *w* = 6, 7, 15, and 22 are just one iteration short of maximal complexity. The missing pattern in those four cases is a *w*-length sequence of all DOWN movements. When the sequence has no UP movements at all, the investor never decides to buy, regardless of the value of *w*.

Figure 4 graphs the various price series from twelve of the thirteen lookback values producing complexity. (The graph of lookback 22 is excluded due to its long period, but it displays a path similar to that of lookback 19.)

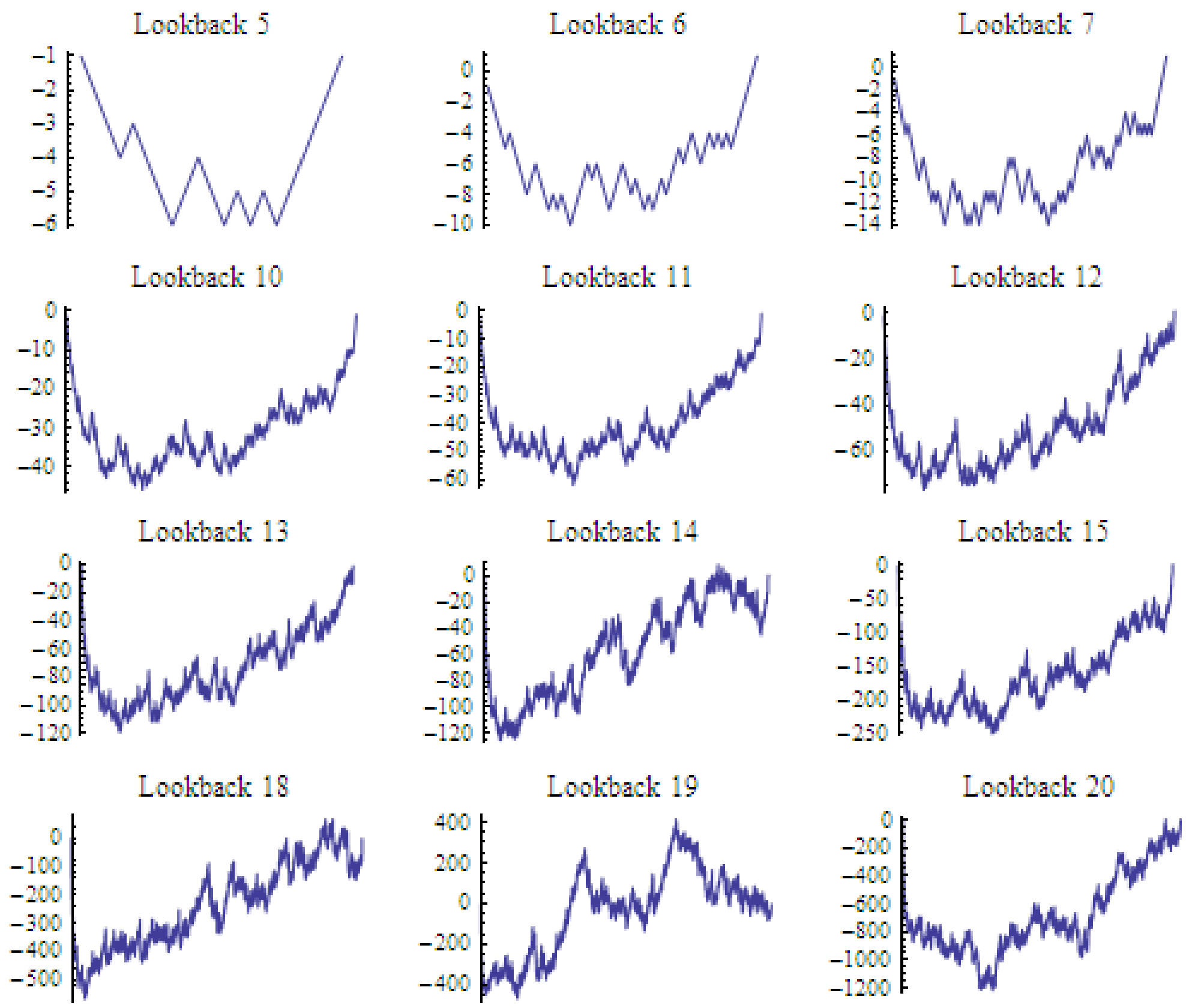

**Figure 4: Graphs of the Complex Prices Series.** The figure below graphs the price series from most of the lookback windows that generate complexity. Each price series is plotted for the length of its period: the next prices in the series will repeat from the beginning. All the graphs begin at zero and all experience sharp drops before recovering. This is because the initial history is a sequence of UPs, which immediately generates, by rule 54, an equal-length series of DOWNs, before becoming more complex.

Note that each of the price series end right near where they begin, at zero. The reason for this is that the periods are very close to the maximum periods. If they exactly equal the maximum period, then the price series traverses each possible sequence of UPs and DOWNs, so for each sequence with more UPs than DOWNs, the exact opposite will also occur at some point, thus cancelling out. When the periods are less than maximal, the final point can differ somewhat from the beginning, though in practice the difference is small.

Notice the frequent busts, mainly at the beginning of the series. That is a consequence of starting with an initial sequence of all UP values. By the logic described above for rule 54, a sequence of UPs would immediately generate an equal-length sequence of DOWNs before becoming more complex. In principle one could start with any initial sequence, which can be visualized just by imagining the price series as beginning later in figure 4 before looping back.

Visually the price series appear similar to those one might see from financial securities, but different from one might expect from a simple random walk or normal distribution. In addition to the large jumps down, there also appear to be fairly frequent jumps of more moderate size both up and down that still appear larger than expected from a normal distribution.

Figure 5 graphs the histograms of the nine largest lookback windows for the moving average of changes. The negative skewness and fat tails appear here as well.

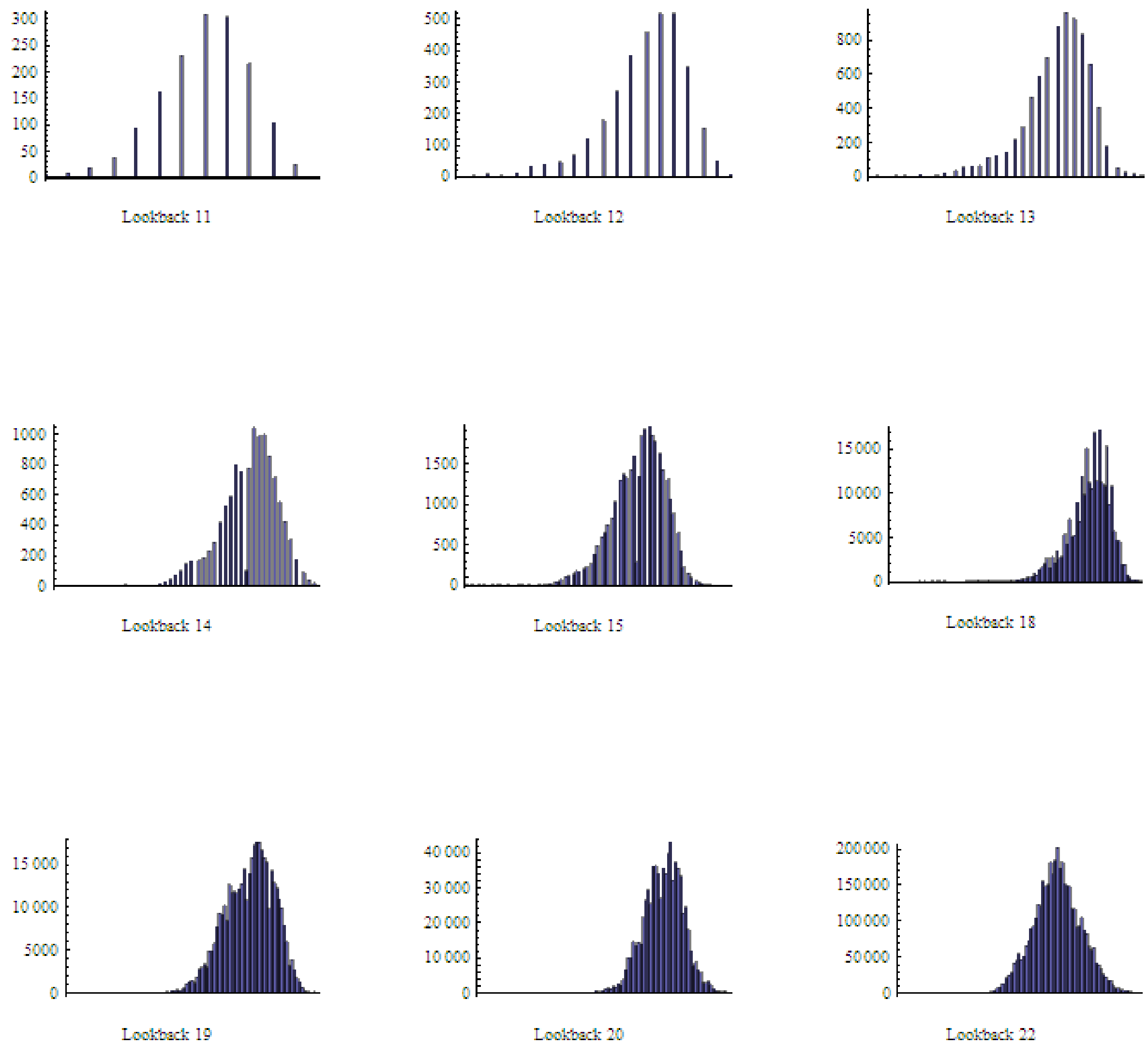

**Figure 5: Histograms of Moving Averages.** For each of the labeled lookback windows *w*, the 100-bar histograms display the average frequencies of the $2^{w-7}$ moving averages of the daily changes generated by rule 54 (so that there are a variety of values instead of just -1 and +1). Notice the negative skewness and fat tails. The white space to the left of the main bulk of the histograms in the bottom row is not empty but rather has rare but extreme down jumps.

We can statistically examine this visual effect. Table 2 lists the skewness and excess kurtosis for the daily distribution of rule 54 with a $w = 22$ lookback window, where a day is defined as a sequence of a given number of ticks. It also lists the standard errors. Following Turner and Weigel (1992), the standard error of skewness is $6/d$ and the standard error of excess kurtosis is $24/d$, where $d$ is the number of days on which the estimate is based.

**Table 2: Summary Statistics of Rule 54 for a 22-tick Lookback.** The table below lists the summary statistics for rolling days generated by rule 54 for a lookback window of 22 ticks, where the deemed length of each day varies from 32 ticks (implying each tick lasts about 15 minutes in a typical 6.5-hour trading day) to 131,072 ticks (implying each tick lasts about 0.18 seconds). The product of the number of ticks per day and the number of days is always $2^{22}$ = 4,194,304. Because each tick is either +1 or -1, the number of distinct *daily* values is always lower than both the number of ticks per day and the number of days. The skewness and excess kurtosis are reported along with their standard errors. Each of the daily distributions are negatively skewed and have fatter tails than a normal distribution.

| Ticks/Day | # Days | Distinct | Skewness | Excess Kurtosis |
|---|---|---|---|---|
| 32 | 131,072 | 26 | -0.34 (0.00005) | 0.19 (0.00018) |
| 64 | 65,536 | 37 | -0.74 (0.00009) | 0.87 (0.00037) |
| 128 | 32,768 | 59 | -0.98 (0.00018) | 2.15 (0.00073) |
| 256 | 16,384 | 75 | -1.08 (0.00037) | 3.06 (0.00146) |
| 512 | 8,192 | 92 | -1.07 (0.0073) | 4.01 (0.00293) |
| 1,024 | 4,096 | 109 | -1.04 (0.00146) | 4.96 (0.00586) |
| 2,048 | 2,048 | 132 | -1.27 (0.00293) | 8.63 (0.01172) |
| 4,096 | 1,024 | 157 | -1.48 (0.00586) | 10.46 (0.02344) |
| 8,192 | 512 | 170 | -1.70 (0.01172) | 11.81 (0.04688) |
| 16,384 | 256 | 154 | -1.73 (0.02344) | 11.20 (0.009375) |
| 32,768 | 128 | 109 | -1.54 (0.04688) | 7.93 (0.18750) |
| 65,536 | 64 | 59 | -2.02 (0.09375) | 8.61 (0.375000) |
| 131,072 | 32 | 30 | -1.65 (0.18750) | 5.03 (0.75000) |

Compared to the normal distribution of a random walk, the minimal model has fatter tails and more frequent crashes. Each row in table 2 has significant negative skewness and excess kurtosis.

Market returns in real life are commonly thought to have negative skewness and fat tails as well (e.g. Turner and Weigel 1992), so the minimal model provides a better empirical match than the simple random walk model.

Note also that there have been only two parameters: the rule number 54 corresponding to the minimal model and the lookback window. The rule number is not a parameter that can be tweaked to give slightly different skewness or kurtosis: all other values for 2-state, buy-sell rules generate non-complex time series. Similarly, the skewness and kurtosis results seem to hold for nearly all values of the lookback window, so long as they generate complexity at all, so the lookback window is also not a parameter that can be modified to, for example, obtain a complex result without negative skewness or excess kurtosis.

Therefore, it appears that, for 2-state, buy-sell rules, if complexity is generated at all, it is generated in a way that seems to match reality better than a random walk.

To be sure, other statistical models such as fat-tailed random walks can have their parameters fit to better match reality as well, but the surprising result is that the minimal model of complexity introduced here is not simply equivalent to the standard random walk, and indeed shows distinctive features common to real markets, without any fitting.

## 5. Conclusion

Financial time series look random and complex. Usually they are modeled with random and complex dynamics such as stochastic processes or interacting traders. But it turns out that a simple model can generate realistic looking series as well. In particular, there exists an essentially unique, simple, and robust model that can be considered the minimal model of financial complexity, which requires only a single investor trading a single market asset.

The minimal model has an interesting interpretation of what the representative investor does: looking at just consecutive ticks, he sells whenever the asset exhibits momentum (two consecutive upticks or two consecutive downticks)\ and buys whenever the asset exhibits mean reversion (an uptick followed by a downtick, or vice versa). Critically, his decision, or its implementation, is delayed by a few ticks. Even delays as short as a dozen or so ticks can generate long-lasting complex price series.

Serendipitously, the minimal model seems to generate more realistic series than those of a random walk, with fatter tails and more frequent crashes, even though it has no parameters that can be tweaked for that purpose.

The minimal model generalizes easily to allow complex variety and can also be generalized to allow multiple traders and multiple assets. Those are the subjects of future research.

## Acknowledgements

I thank Stephen Wolfram and Jason Cawley for many helpful comments during the 2007 NKS Summer School and two anonymous referees for additional helpful suggestions.